\documentclass[symmetry,article,submit,XeLaTeX,moreauthors]{Definitions/mdpi}

\firstpage{1}
\pubvolume{0}
\issuenum{1}
\articlenumber{0}
\pubyear{2025}
\copyrightyear{2025}
\datereceived{ }
\daterevised{ } 
\dateaccepted{ }
\datepublished{ }
\preto{\abstractkeywords}{\nolinenumbers}
\hreflink{https://doi.org/} 

\Title{The Principle of Maximum Conformality Correctly Resolves the Renormalization-Scheme-Dependence Problem}

\TitleCitation{The Principle of Maximum Conformality Correctly Resolves the Renormalization-Scheme-Dependence Problem}

\Author{Jiang Yan $^{1}$\orcidA{}, Stanley J. Brodsky $^{2}$\orcidB{}, Leonardo Di Giustino $^{3,4}$\orcidC{}, Philip G. Ratcliffe $^{3,4}$\orcidD{}, Sheng-Quan Wang $^{5}$ and Xing-Gang Wu $^{1}$*\orcidE{}}

\AuthorNames{Jiang Yan, Stanley J. Brodsky, Leonardo Di Giustino, Philip G. Ratcliffe, Sheng-Quan Wang and Xing-Gang Wu}

\isAPAStyle{%
	\AuthorCitation{Lastname, F., Lastname, F., \& Lastname, F.}
}{%
	\isChicagoStyle{%
		\AuthorCitation{Lastname, Firstname, Firstname Lastname, and Firstname Lastname.}
	}{
		\AuthorCitation{Yan, J.; Brodsky, S. J.; Giustino, L. D.; Ratcliffe, P. G.; Wang, S.-Q.; Wu, X.-G.}
	}
}

\address{%
	$^{1}$ \quad Department of Physics, Chongqing University, Chongqing 401331, P.R. China; yjiang@cqu.edu.cn~(J. Yan); wuxg@cqu.edu.cn~(X.-G. Wu)\\
	$^{2}$ \quad SLAC National Accelerator Laboratory, Stanford University, Stanford, California 94039, USA; sjbth@slac.stanford.edu~(S.~J.~Brodsky)\\
	$^{3}$ \quad Department of Science and High Technology, University of Insubria, via Valleggio 11, I-22100 Como, Italy\\
	$^{4}$ \quad INFN, Sezione di Milano--Bicocca, piazza della Scienza 3, I-20126 Milano, Italy; leonardo.digiustino@uninsubria.it~(L.~Di~Giustino); philip.ratcliffe@uninsubria.it~(P.~G.~Ratcliffe)\\
	$^{5}$ \quad Department of Physics, Guizhou Minzu University, Guiyang 550025, P.R. China; sqwang@alu.cqu.edu.cn~(S.-Q. Wang)\\}

\corres{Correspondence: wuxg@cqu.edu.cn}

\abstract{In this paper, we clarify a serious misinterpretation and consequent misuse of the Principle of Maximum Conformality (PMC), which also can be served as a mini review of PMC. In a recently published article~\cite{Stevenson:2023isz}, P.~M. Stevenson has claimed that ``the PMC is ineffective and does nothing to resolve the renormalization-scheme-dependence problem", concluding incorrectly that the success of PMC predictions is due to the PMC being a ``laborious, \emph{ad hoc}, and back-door" version of the Principle of Minimal Sensitivity~(PMS). We show that such conclusions are incorrect, deriving from a misinterpretation of the PMC and overestimation of the applicability of the PMS. The purpose of the PMC is to achieve precise fixed-order pQCD predictions, free from conventional renormalization scheme and scale ambiguities. We demonstrate that the PMC predictions satisfy all of the self-consistency conditions of the renormalization group and standard renormalization-group invariance; the PMC predictions are thus independent of any initial choice of renormalization scheme and scale. The scheme independence of the PMC is also ensured by commensurate scale relations which relate different observables to each other. Moreover, in the Abelian limit, the PMC dovetails into the well-known Gell-Mann--Low framework, a method universally revered for its precision in QED calculations. Due to the elimination of factorially-divergent renormalon terms, the PMC series not only attains a convergence behavior far superior to that of its conventional counterparts but also deftly curtails any residual scale dependence caused by the unknown higher-order terms. This refined convergence, coupled with its robust suppression of residual uncertainties, furnishes a sound and reliable foundation for estimating the contributions from unknown higher-order terms. Anchored in the bedrock of standard renormalization group invariance, the PMC simultaneously eradicates the factorial divergences and eliminates superfluous systematic errors, which inversely provides a good foundation for achieving high-precision pQCD predictions. Consequently, owing to its rigorous theoretical underpinnings, the PMC is eminently applicable to virtually all high-energy hadronic processes.}


\begin{document}

P.~M. Stevenson has recently claimed~\cite{Stevenson:2023isz} that the Principle of Maximum Conformality (PMC)~\cite{Brodsky:2011ta, Brodsky:2011ig, Brodsky:2012sz, Brodsky:2012rj, Mojaza:2012mf, Brodsky:2013vpa}, developed by Brodsky \emph{et al}., ``is ineffective and does not solve the renormalization-scheme-dependence problem.'' His argument is not built on new evidence but rather on a reassertion of the Principle of Minimal Sensitivity (PMS)~\cite{Stevenson:1980du, Stevenson:1981vj}, which he himself proposed in 1980. Moreover, he leans heavily on his earlier 1983 criticism~\cite{Celmaster:1982zj} of the effectiveness of the method of Brodsky--Lepage--Mackenzie~(BLM) proposed in 1982~\cite{Brodsky:1982gc}. Back then, the vigorous debate between the PMS and the BLM aroused people's great interests over the QCD scale-setting problem, a discussion that significantly advanced both the theoretical understanding and the precision of perturbative QCD (pQCD).

The QCD Lagrangian for massless quarks is invariant under conformal transformations. This symmetry is broken by quantum corrections due to the fact that the conformal symmetry of a quantum field theory implies that the $\beta$-function must vanish~\cite{Braun:2003rp}. Stevenson states in Appendix~B of Ref.~\cite{Stevenson:2023isz} that one can choose a proper/optimal renormalization scheme to achieve exact conformality; this statement is incorrect. Stevenson's suggested procedure, the PMS, mandates that the optimal renormalization scheme and scale be identified by forcing the derivative of any given fixed-order series with respect to renormalization scheme and scale choices to vanish. In doing so, it tacitly presumes that all contributions from uncalculated higher-order terms are null, which is a unfounded assumption that contravenes standard renormalization group invariance~(RGI)~\cite{Wu:2014iba}~\footnote{Stevenson states that none of the points listed in Page~2 of Ref.~\cite{Stevenson:2023isz}, especially Eqs.~(8) and (11) there, are ``\emph{in any way dependent upon the PMS}." This is, in fact, a disguised replacement of concept, since those points are surely correct for any physical observable that corresponds to an infinite-order series.

For simplicity, one can consider applying the PMS to a fixed-order series, Stevenson's Eqs.~(8) and (11) are the PMS basic approximations for fixed-order series.}. In essence, the PMS procedure is nothing more than finding the extreme point of a given fixed-order series. Since the idea is transparent, the further developments of the PMS method are then mainly concentrated on how to improve the quantitative analysis with better numerical precision~\cite{Stevenson:1985dy, Stevenson:2012ti, Ma:2014oba, Ma:2017xef}. This explains why Stevenson's comments on BLM/PMC are still only based on early arguments at that time, even though many years have since passed. Moreover, his successive comments and assumed ``PMC samples" given in Ref.~\cite{Stevenson:2023qsh} are unfortunately full of typos and wrong deductions, indicating he does not understand the PMC at all. To show how to apply the PMC correctly, we have put a detailed explanation of his misuse in the Appendix~\ref{app}.

On the other hand, there have been a number of important developments of the BLM method. In 1984 and 1992, Grunberg observed that the above criticism can be softened by using the method of effective charges~\cite{Grunberg:1984py, Grunberg:1992mp}. In 1994, Brodsky \emph{et al}.\ found that the criticism can be clarified by using commensurate scale relations (CSRs) among effective charges~\cite{Brodsky:1994eh}. The CSRs can relate the pQCD approximants of physical observables to each other, which then ensure that the BLM pQCD predictions are independent of the choice of renormalization scheme. The interested reader may consult Ref.~\cite{Brodsky:1994eh} for a detailed explanation of scheme independence in the BLM predictions~\footnote{Ref.~\cite{Brodsky:1994eh} gives an explanation of leading-order CSRs, and a novel demonstration of the scheme independence of CSRs to all orders using the PMC language has been finished in 2021~\cite{Huang:2020gic}.}. The transitivity and symmetry properties of the commensurate scales are the scale transformations of the renormalization ``group" as originally defined by Stueckelberg and Petermann~\cite{StueckelbergdeBreidenbach:1952pwl, Peterman:1978tb}. Thus, Stevenson's claim that ``the BLM/PMC does nothing to resolve the renormalization-scheme-dependence problem" was already clarified in the BLM language more than twenty years ago. This observation was more recently confirmed in 2012, e.g.\ Ref.~\cite{Brodsky:2012ms} which presents a demonstration that the BLM/PMC predictions satisfy all the RG self-consistency conditions, such as reflectivity, symmetry and transitivity. This reference also demonstrates that, in contrast, the PMS predictions do not satisfy those self-consistency conditions, explicitly transitivity, so that the PMS relations between observables depend on the choice of intermediate renormalization scheme.

It should be emphasized that the goal of resolving the conventional scale-setting ambiguity is not to find the optimal scheme and/or optimal scale of the initial pQCD series, but to achieve an improved pQCD series that is free of any choice of scheme and scale. The BLM/PMC achieves this goal.

The BLM method automatically resums both the color-octet gluon exchange and the quark-pair vacuum-polarization contributions into the running behavior of $\alpha_s$, which is a function of $\beta_0=11{C_A}-{2\over 3} n_f$, at leading order, where $C_A=3$ is the color factor and $n_f$ is the number of active quark flavors. The BLM method has achieved many successes and has so far been cited over 1200 times\,\footnote{The PMS paper~\cite{Stevenson:1981vj} has also been cited over 1200 times, although most of the citations are not for the PMS method itself, but for the suggested method of extended RGEs, which provides a convenient way of analyzing the scheme and scale running behavior simultaneously.}. However, when one calculates the QCD corrections at higher than next-to-leading order (NLO), the problem of correctly resumming the $n_f$ power series into the coupling $\alpha_s$ is encountered.

Since the BLM/PMC method works well, it has aroused great interest in understanding the underlying principles and in finding correct methods for extending it to all orders. Many attempts have been tried in the literature, which include the dressed skeleton expansion, the large $\beta_0$-expansion, the BLM expansion with an overall effective scale, the sequential BLM etc.~\cite{Lu:1991yu, Grunberg:1991ac, Beneke:1994qe, Neubert:1994vb, Lovett-Turner:1995zwc, Ball:1995ni, Brodsky:1995tb, Brodsky:1997vq, Braaten:1998au, Hornbostel:2002af, Mikhailov:2004iq}. The emphasis of most of those references is just to eliminate the $n_f$-terms or to improve the pQCD convergence; however, such procedures to extend BLM do not simultaneously satisfy all the RG self-consistency conditions. Thus, although these early methods have lead to some improvements, the criticisms of BLM made in some references are incorrect since the problems are actually due to improper extension of BLM to higher orders. In fact, such methods do not retain its most important feature: namely, that the BLM prediction should be independent of any choice of renormalization scheme and scale.

In 2011 the BLM procedure was extended into an all-orders method of PMC~\cite{Brodsky:2011ta, Brodsky:2011ig}, the purpose of which is to simultaneously solve both the conventional renormalization scheme and scale ambiguities. The PMC is stimulated by the observation that if good matching of the expansion coefficients with the corresponding coupling $\alpha_s$ can be achieved, then exactly scheme-independent predictions can be obtained. This demonstrates that dealing with the $\{\beta_i\}$-terms involved in the renormalization group equation (RGE), which control the breaking of conformality in the perturbative series, is more fundamental than dealing with the $n_f$-terms alone~\cite{Brodsky:1994eh}. The PMC thus perfects essential features of BLM.

In 2012, ``degeneracy relations" among different orders of a pQCD series were suggested; these relations ensure that the transformation of the $n_f$-terms into $\{\beta_i\}$-terms is made in a unique way~\cite{Mojaza:2012mf, Brodsky:2013vpa}. In 2015, we demonstrated that such degeneracy relations are general properties of QCD theory~\cite{Bi:2015wea}\,\footnote{Recently, we have derived new degeneracy relations with the help of the RGEs involving both the $\beta$-function and the quark mass anomalous dimension $\gamma_m$-function~\cite{Huang:2022rij}, which leads to an alternative PMC scale-setting procedure that simultaneously fixes the correct magnitudes of the $\alpha_s$ and the $\overline{\rm MS}$-scheme quark mass of the perturbative series. The characteristic operator has been suggested to formalize those PMC procedures~\cite{Yan:2024oyb}. }.

The PMC determines the magnitude of $\alpha_s$ by using the RGE; its arguments (the so-called PMC scales) are obtained by shifting the initial argument of $\alpha_s$ to eliminate all the RGE-involved non-conformal $\{\beta_i\}$-terms. The PMC scales thus reflect the virtuality of the propagating gluons for the QCD processes. The resulting perturbative series then matches the scheme-independent conformal series with $\beta=0$. The PMC inherits all the good features of BLM. For example: 1) In the Abelian limit the PMC reduces to the Gell-Mann--Low~(GML) method~\cite{Gell-Mann:1954yli} -- this analytic limit provides an important constraint on the renormalization scale-setting problem in QCD. In contrast, the PMS cannot satisfy this limit, and it cannot reproduce the GML scale for QED observables, 2) The PMC achieves scheme- and scale-independent pQCD predictions, and 3) Because of the elimination of the factorially-divergent renormalon terms, which could be $\alpha_s^n \beta^n_0 n !$ by using large $\beta_0$-approximation~\cite{Beneke:1994qe, Neubert:1994vb}, the PMC series naturally becomes more convergent\,\footnote{There are cases which accidentally have large cancellations among the conformal and non-conformal coefficients for a specific choice of scale under a conventional scale-setting approach and the elimination of the divergent renormalon terms alone may lead to an even weaker convergence than conventional series. This however does not mean that conventional series is better than the PMC one; since the PMC series is scale-invariant and represents the intrinsic perturbative nature of the series, whereas the conventional series is largely scale dependent and the large cancellation for a specific scale usually disappears on choosing another scale.}. In fact, we have shown that the PMC provides a systematic way to extend BLM scale-setting from the NLO-level to all-orders; see the reviews~\cite{Wu:2013ei, Wu:2019mky, DiGiustino:2023jiq}, in which many successful examples are also given.

The main argument that the PMC ``does not work" lies below Eq.~(4) of Ref.~\cite{Stevenson:2023isz}. Stevenson's Eq.~(4) itself is correct, showing that the RGE determines the scale-running behavior of $\alpha_s$. This RGE is surely scheme dependent. He then states that, since the PMC is a scale-setting approach, it cannot solve the scheme-fixing problem, and thus cannot work. However, this argument is incorrect, being based on a misunderstanding of the PMC. In fact, if the scheme-dependent RGE is correctly applied to the pQCD series to set the correct magnitude of $\alpha_s$, then a scale-invariant conformal series is obtained.  Since the conformal coefficients are well matched to the corresponding $\alpha_s$ coefficients at each order, the resulting PMC series will be simultaneously scheme-independent. In fact, the above-mentioned CSRs among different PMC pQCD approximants also ensure the scheme independence of the PMC predictions. Thus, the PMC does indeed solve the scheme and scale ambiguities simultaneously.

We remark that the PMC uses the RGE to reabsorb the $n_f$-terms specifically related to the UV-divergent diagrams. The PMC procedure works with \emph{any} initial scheme definition. Stevenson claims his Eq.~(7) to be valid in general, but this also includes variations of the color parameters. This cannot be correct, since a change in the color structure corresponds to a change in the gauge structure of the initial SU(N) theory, which ends up mixing results from different theories. Introducing terms that modify the color structure of the $\beta_0$ and $\beta_1$ coefficients implies that the finite and divergent parts of the UV-divergent diagrams run with different $\{\beta_i\}$ terms. This is not correct and leads to incorrect results: the $\{\beta_i\}$ coefficients entering the RGE must be identical to those defined by the renormalization scheme. Thus, Stevenson's  Eq.~(16) is incorrect.

The work of Banks and Zaks \cite{Banks:1981nn} directly contradicts a further statement by Stevenson: the QCD strong coupling develops a conformal window in the interval $\frac{34N_c^3}{13N_c^2-3}<N_f<\frac{11}{2}N_c$ and has a non-interacting fixed point at $N_f=\frac{33}{2}$~\cite{Banks:1981nn, Gardi:1998qr, DiGiustino:2021nep, DiGiustino:2023jiq}, which corresponds to the asymptotically free limit of QCD. Stevenson then fully contradicts himself when he states that the $\mathcal{C}(\delta)$ in Eq.~(6) is scheme invariant since it depends only on $\beta_0$ and $\beta_1$, but then in his Eqs.~(20) and (21) he shows that it is scheme dependent. Indeed, we hold both the $\mathcal{C}(\delta)$ and the $C_1^*$ conformal coefficient to be scheme invariant according to the ``proper" extended RGE transformations. We refer to ``proper" scheme transformations as those that may be achieved by an RGE transformation. Thus, any scheme change translates into a scale transformation leaving the $\mathcal{C}(\delta)$ formally invariant. This corresponds to having the right-hand-side of Eq.~(7) of Ref.~\cite{Stevenson:2023isz} void of any dependence on color factors.

Moreover, if the color structure for the coupling is altered, then for self-consistency, the structure of the entire fixed-order calculation must also be varied accordingly; however, one cannot adopt such a procedure, since a change in the color structure corresponds to a change of the initial SU(N) theory (e.g.\ from QCD to QED, see Ref.~\cite{Brodsky:1997jk}). Thus, any relation among couplings in different schemes, which may have perturbative validity in QCD, but which cannot be considered as ``proper" extended RGE transformations, should be considered as matching relations among quantities defined in different approximations or obtained using different approaches. In fact, starting from a given exact theory and using different approximations or approaches, different ``scheme" definitions may be obtained. This is only a matter of using different strategies for calculations. However, in this case too, results can be improved by using the PMC, and the residual dependence on the particular implicit definition of the ``scheme" can be suppressed perturbatively by including higher-order calculations.

It can also be explicitly shown that the results obtained using PMC are scheme independent (see e.g., Ref.~\cite{Wu:2018cmb}). As an important example, the scheme independence of the generalized Crewther relation provides a fundamental relation between the non-singlet Adler function and the Bjorken sum rules for polarized deep-inelastic electron scattering~\cite{Bjorken:1967px, Bjorken:1969mm, Broadhurst:1993ru, Brodsky:1995tb, Crewther:1997ux, Gabadadze:1995ei, Braun:2003rp}; this example confirms that the scheme independence can be achieved by using the PMC~\cite{Shen:2016dnq, Huang:2020gic}.

We emphasize that, in order to apply the PMC correctly, it is important to distinguish the nature of different $n_f$-terms; i.e., whether they are related to ultraviolet-finite contributions (such as light-by-light scattering in QED), or to  the running of the QCD coupling, or to the running of quark masses, and, in a deeper analysis, to particular UV-divergent diagrams (as discussed in Ref.~\cite{DiGiustino:2023jiq}). Once all $n_f$-terms have been associated with the correct diagrams or parameters, the conformal coefficients become RGE invariant and match the coefficients of a conformal theory. Moreover, a deep insight into the QCD strong coupling $\alpha_s(Q^2)$ at all scales, including $Q^2=0$, has been recently achieved (see e.g. Refs.~\cite{Deur:2023dzc, Deur:2017cvd}), again showing the results are consistent with the PMC.

The PMC provides a consistent prescription for scale fixing by reabsorbing all scheme-dependent terms of, say, a cross-section into the running coupling and the PMC scale; this leads to the minimization/cancellation of the effects of scale and scheme uncertainties in the perturbative series. It should be emphasized that the CSRs between the pQCD approximates are impervious to the choice of renormalization scheme.

We now comment on Stevenson's second point of view on the PMC predictions: namely, that the reason ``why the PMC predictions often seem quite good" is because the PMC introduces ``residual scheme dependence" and has to introduce some way to suppress it, thus rendering it a back-door imitation of the PMS. These deductions on ``maximum conformality" are simply not valid, being based on an early misunderstanding of BLM/PMC and completely disregarding recent developments in BLM/PMC. Principally, as explained above, as the successor of BLM, the PMC prediction is scheme independent and has no residual scheme dependence. Practically, conventional scheme for defining the running coupling $\alpha_s$ suffers from a complex and scheme-dependent RGE, which is usually solved perturbatively at higher orders owing to the entanglement of its scheme-running and scale-running behaviors. If the complex $\{\beta_i\}$-terms in the higher-order pQCD series are not dealt with precisely\,\footnote{For example, not all $n_f$-terms should be transformed as the $\{\beta_i\}$-terms; if the $n_f$-terms are from the UV-free light-by-light diagrams, they should be treated as conformal coefficients and be unchanged when applying the PMC.}, these complications may lead to residual scheme dependence even after applying the PMC. This can however be avoided by using the $C$-scheme coupling $\hat\alpha_s$ suggested in 2016~\cite{Boito:2016pwf}, whose scheme-running and scale-running behaviors are governed by the same scheme-independent RGE. Consequently, one can achieve an analytic solution for $\alpha_s$ at any fixed order. In fact, by using the $C$-scheme coupling together with the PMC single-scale-setting approach~\cite{Shen:2017pdu}, a demonstration that the PMC prediction is scheme-independent to all orders for any renormalization scheme was provided in 2018~\cite{Wu:2018cmb}.

There are, of course, uncertainties related to the unknown higher-order terms. In the case of any perturbative series, for example, there are two kinds of residual scale dependences for the PMC series~\cite{Zheng:2013uja} due to uncalculated higher-order contributions. The PMC scale itself has a perturbative expansion in $\alpha_s$, this leads to the first kind of residual scale dependence for the PMC scale. In addition, the last terms of the pQCD approximant are unfixed, since its magnitude cannot be determined; this is the second kind of residual scale dependence. These residual scale dependences are distinct from the conventional scale ambiguities and are suppressed due to the perturbative nature of the PMC scale.

The PMC ``single-scale" method was introduced in 2017~\cite{Shen:2017pdu}  in order to suppress the residual scale dependence; this method also renders the PMC scale-setting procedures simpler and more easily automated, The single effective PMC scale is determined by requiring all the RGE-involved non-conformal terms to vanish simultaneously; it can be regarded as the overall effective momentum flow of the process. The PMC single-scale method exactly removes the second kind of residual scale dependence. The overall effective PMC scale displays stability and convergence with increasing order in pQCD, which can approach the precision of N$^{n-1}$LL-accuracy for a N$^{n}$LO pQCD series when the power of $\alpha_s$ of its leading-order terms is $\geq 1$; and the first kind of residual scale dependence is thus highly suppressed.

In addition to eliminating the renormalization-scheme and -scale ambiguities at fixed-order, the predictive power of pQCD depends on the important issue of finding a reliable way to estimate the magnitudes of unknown higher-order terms using information from the known pQCD series. The PMS cannot do this self-consistently, since the contributions of the unknown terms are set to zero at the first stage. In contrast, after applying the PMC, the scheme- and scale-invariant series provides a precise basis for estimating the unknown contributions.  In 2020, we uncovered an additional property of renormalizable SU(N)/U(1) gauge theories~\cite{DiGiustino:2020fbk}, called  ``Intrinsic Conformality (iCF)", which underlies the scale invariance of physical observables. This property demonstrates that the scale-invariant perturbative series displays the intrinsic perturbative nature of a pQCD observable. In 2022, following the idea of iCF, we then suggested another novel single-scale setting approach which utilizes the PMC~\cite{Yan:2022foz} and also proved the two PMC single-scale methods to be equivalent. This equivalence indicates that using the RGE to fix the value of the effective coupling is equivalent to requiring each loop term to satisfy scale invariance simultaneously, and vice versa.

In 2018~\cite{Du:2018dma} and 2022~\cite{Shen:2022nyr}, it was found that, using the Pad\'{e}-    approximation and Bayesian approaches accordingly, the PMC series provides a reliable basis for obtaining constraints on the predictions for the uncalculated higher-order contributions, thus extending the predictive power of pQCD.

In summary, we have shown that Stevenson's criticisms of the PMC are unfounded, being based only on incorrect features of the PMS, as well as a disregard of the considerable achievements of BLM/PMC which have been developed over recent years.

It has been shown that the purpose of the PMS is to determine an optimal scheme and optimal scale for a given pQCD series. To achieve the goal, the PMS assumes that all uncalculated higher-order terms contribute zero, which then fixes the desired optimal values by mathematically requiring the partial derivatives of the pQCD series with respect to scheme and scale choices to vanish. The PMS method does provide a scheme-independent estimate around the determined optimal point, but it violates the symmetry and transitivity properties of the RG~\cite{Brodsky:2012ms} and does not even reproduce the GML scale for QED observables. The PMS predictions thus have serious flaws and weak points. The convergence of the PMS series is often questionable, which does not agree with the usual perturbative behavior of the series, even worse than that of conventional series. This explains why the PMS cannot offer correct lower-order predictions~\cite{Ma:2014oba}. Moreover, the PMS cannot achieve a reliable prediction on the contributions of uncalculated higher-order terms. As an explicit example, in 1987 and 1990, Kramer and Lampe~\cite{Kramer:1987dd, Kramer:1990zt} analyzed the jet production fractions in $e^+e^-$ annihilation, showing that the optimal PMS scale increases without bound at small energy scales of the jet fraction, which indicates that the PMS scale do not have the correct physical behavior in the limit of small jet energy.

On the other hand, the PMC method provides a systematic way of eliminating conventional renormalization scheme and scale ambiguities; it has a rigorous theoretical foundation, satisfying standard RGI and all self-consistency conditions derived from the RG.  We emphasize that the CSRs between physical observables ensure the PMC predictions to be independent of the choice of renormalization scheme.

The PMC scales are obtained by shifting the argument of $\alpha_s$ to eliminate all non-conformal $\{\beta_i\}$-terms; the PMC scales thus reflect the  physical virtuality of the propagating gluons in the QCD process. The factorially-divergent renormalon contributions are eliminated, as they are summed into $\alpha_s$, and thus the resulting pQCD convergence is in general greatly improved. In the Abelian limit, the PMC method reduces to the GML method for precision tests of QED~\cite{Brodsky:1997jk}. The resulting scale invariant and convergent PMC method retains the underlying principles and features of the BLM method, extending it unambiguously to all orders. The resulting conformal series obtained by applying the PMC is scheme and scale independent. The resulting scale-invariant and convergent PMC series thus also provides a reliable basis for obtaining constraints on uncalculated higher-order contributions, greatly extending  the predictive power of pQCD.  The PMC thus improves precision tests of the Standard Model and increases the sensitivity of experiments to new physics beyond the standard model.

\hspace{1cm}

\acknowledgments{Project supported by graduate research and innovation foundation of Chongqing, China (Grant No.~CYB240057), the Natural Science Foundation of China under Grant Nos.~12175025, 12265011, and 12347101, and by the Department of Energy Contract No.~DE-AC02-76SF00515. SLAC-PUB-17755.}

\abbreviations{Abbreviations}{
	The following abbreviations are used in this manuscript:\\
	
	\noindent
	\begin{tabular}{@{}ll}
		PMC & Principle of Maximum Conformality\\
		PMS & Principle of Minimal Sensitivity\\
		QED & quantum electrodynamics\\
		QCD & quantum chromodynamics\\
		pQCD & perturbative quantum chromodynamics\\
		BLM & Brodsky--Lepage--Mackenzie\\
		GML & Gell-Mann--Low\\
		RG & renormalization group\\
		RGE & renormalization group equation\\
		RGI & renormalization group invariance\\
		CSR & commensurate scale relation\\
		NLO & next-to-leading order\\
		UV & ultraviolet\\
		iCF & Intrinsic Conformality
	\end{tabular}
}

\appendixtitles{yes} 
\appendixstart
\appendix
\section[\appendixname~\thesection]{Explanations on the Stevenson's misuse of PMC}\label{app}

In his article~\cite{Stevenson:2023qsh}, Stevenson continues his arguments of Ref.~\cite{Stevenson:2023isz} on the BLM/PMC and presents two extreme examples to support his viewpoints. Since BLM has developed into PMC, we will use PMC to replace BLM/PMC throughout this Appendix. We observe that Stevenson misuses the BLM/PMC scale-setting procedures in these two examples, thus leading to incorrect PMC predictions. Thus, all of his comments on PMC are incorrect. In the following, we will first give the correct PMC procedures and then explain why we say his PMC predictions are wrong. Especially, we show that if the PMC scale-setting procedures have been correctly used, one can achieve scheme-independent PMC predictions, well satisfying the required RGI.

Following the notation of Refs.~\cite{Stevenson:2023isz, Stevenson:2023qsh}, a pQCD series (for simplicity but without loss of generality, we consider the case where the LO $\alpha_s$-power is $1$) is given as
\begin{align}\label{rho0}
	\rho &= c_{0} a + c_{1} a^{2} +\cdots,
\end{align}
where $a=\alpha_s/\pi$ and the $\alpha_s$ running behavior is governed by the RGE or $\beta$-function,
\begin{align}
	\beta(a) \equiv \frac{{\rm d}a}{{\rm d}\ln \mu} = -ba^{2}(1 + c a + \cdots),
\end{align}
with
\begin{align}
	b = \frac{1}{2}\left(11-\frac{2}{3}n_{f}\right),\quad c= \frac{1}{8b}\left(102-\frac{38}{3}n_{f}\right),
\end{align}
and the value of the NLO running-coupling $a$ at scale $\mu$ is given by the root of
\begin{align}\label{RGEsolution}
	\frac{1}{a}\left(1+ca\ln\left|\frac{ca}{1+ca}\right|\right)=b\ln\left(\mu/\tilde{\Lambda}\right),
\end{align}
where $\tilde{\Lambda}$ is the free QCD parameter.

\subsection{Brief introduction of PMC procedures}

For clarity, we briefly introduce the main steps of the PMC method for dealing with the NLO pQCD series (\ref{rho0}), which can be rewritten as
\begin{align}\label{rho}
	\rho &= c_{0} a + (B + A\; n_{f}) a^{2} +\cdots.
\end{align}
We reiterate that the coefficient $A$ of the $n_{f}$-term must be associated with the RGE. There may be other $n_f$-terms which are not related to the running behavior of $\alpha_s$; such $n_f$ terms must be treated as conformal terms, and they thus must be incorporated into the parameter $B$~\cite{Brodsky:2013vpa, Brodsky:1994eh}.

Then we can use the RGE to fix the correct magnitude of $\alpha_s$ for the considered pQCD series. For this purpose, we rewrite Eq.~\eqref{rho} in the following form with the help of the general QCD degeneracy relations~\cite{Bi:2015wea}, e.g.\
\begin{align}\label{rhodegen}
 	\rho = c_{0,0} a + \left(c_{1,0}+c_{1,1}b\right) a^{2} +\cdots.
\end{align}
By absorbing the non-conformal term $c_{1,1} b$ into a new effective coupling $a_{*}$ that is defined via the expression~(i.e. Eq.~(A.10) in Ref.~\cite{Stevenson:2023qsh} or Eq.~(3.4) in Ref.~\cite{DiGiustino:2023jiq})
\begin{align}\label{scale-displacement}
 	a_{*} = a - ba^{2}\ln\left(\mu_{r}^{*}/\mu_{r}\right) +\cdots,
\end{align}
one obtains the following conformal series
\begin{align}\label{rhoPMC}
 	\rho_{\rm PMC} = c_{0,0} a_{*} + c_{1,0} a_{*}^{2} +\cdots.
\end{align}
Comparing Eqs.~(\ref{rhodegen}), (\ref{scale-displacement}) and (\ref{rhoPMC}), we have
\begin{displaymath}
\mu_{r}^{*} = \mu_{r}\exp(-c_{1,1}/c_{0,0}) = \mu_{r}\exp(3A),
\end{displaymath}
and the conformal coefficient $c_{1,0}$ equivalent to $c_{1,0} = c_{1}|_{n_{f}=33/2}$\,\footnote{In Ref.~\cite{Stevenson:2023isz}, Stevenson incorrectly claimed, ``However, making $b$ vanish (by setting $n_{f}=33/2$) does not produce a conformal theory (the $\beta$ function does not vanish but becomes of the form $ha^{3}(1+\cdots)$, where $h=-bc=-(153-19n_{f})/12=107/8$).'' In fact, the NLO conformal coefficient $c_{1,0}=c_{1}|_{n_{f}=33/2}$, but the higher-order conformal coefficients $c_{i>1,0}$ are not simply equal to $c_{i}|_{n_{f}=33/2}$, but instead contain some subtracted terms. For example, $c_{2,0}=c_{2}|_{n_{f}=33/2} - c_{1,1}(bc)|_{n_{f}=33/2}$, where $(bc)$ is the second coefficient of the $\beta$-function. These subtracted terms are designed to make the higher-order coefficients of the $\beta$-function vanish, and the choice of $n_{f}=33/2$ is merely a mathematical treatment. Strictly speaking, the relationship between $c_{2}$ and $c_{2,0}$ arises from the degeneracy relations, that is, $c_{2}=c_{2,0}+2b\,c_{2,1}+b^{2}c_{2,2}+(bc)c_{1,1}$.}. Since $a_{*}$ is fixed by the $\beta$-terms of the process, it represents the effective coupling of the process, and we can obtain its correct magnitude by substituting $\mu_{r}^{*}$ into Eq.~\eqref{RGEsolution}. It should be clarified that Eq.~\eqref{scale-displacement}~(i.e. Eq.~(A.10) in Ref.~\cite{Stevenson:2023qsh} or Eq.~(3.4) Ref.~\cite{DiGiustino:2023jiq}) describes the running behavior of $a$, guiding us on how to incorporate the non-conformal $\beta$-terms into the effective coupling $a_{*}$ to obtain the leading-log accuracy PMC scale $\mu_{r}^{*}$ (to achieve higher accuracy, see Refs.~\cite{Shen:2017pdu, Brodsky:2013vpa}). It however does not allow us to determine the precise value of $a_{*}$, which still needs to be derived by solving the corresponding $\beta$-function (for example, the specific value of the NLO $a_{*}$ should be determined from Eq.~\eqref{RGEsolution}). However, in Ref.~\cite{Stevenson:2023qsh}, Stevenson directly used Eq.~\eqref{scale-displacement} to derive the magnitude of $a_{*}$, which is incorrect.

\subsection{Comments on the Stevenson's first example}\label{firstexample}

The $e^{+}e^{-}$ total cross-section ratio $\mathcal{R}_{e^{+}e^{-}} = \sigma(e^{+}e^{-}\to {\rm hadrons})/\sigma(e^{+}e^{-}\to \mu^{+}\mu^{-}) = 3\sum_{q}e_{q}^{2}\left[1+R(Q)\right]$, where $Q=\sqrt{s}$ is the $e^{+}e^{-}$ center-of-mass collision energy at which the ratio is measured, is known up to $\mathcal{O}(\alpha^4_s)$ in the $\overline{\rm MS}$-scheme~\cite{Baikov:2008jh, Baikov:2010je, Baikov:2012zn, Baikov:2012zm}. The PMC analysis at this order can be found in the review~\cite{Wu:2019mky}. Stevenson's first example is based on the NLO series~\cite{Stevenson:2023qsh}, i.e.
\begin{align}
	R(Q) = r_{0} a + r_{1} a^{2} +\cdots, \label{convRQ}
\end{align}
whose perturbative coefficients $r_{i=\{0,1\}}$ at the scale $\mu_{r}=Q$ in the $\overline{\rm MS}$-scheme are~\cite{Baikov:2012zm}
\begin{align}
	r_{0} &= \frac{3}{4} \gamma_{0}^{\rm NS},\label{r0}\\
	r_{1} &= \frac{3}{4} \gamma_{1}^{\rm NS} + \frac{3}{8} p_{1}^{\rm NS} b(n_{f}),\label{r1}
\end{align}
where $b(n_{f}) = (33-2n_{f})/6$ and
\begin{align}
	\gamma_{0}^{\rm NS} &= C_{F} = \frac{4}{3},\label{gamma0}\\
	\gamma_{1}^{\rm NS} &= -\frac{1}{8}C_{F}^{2}+\frac{133}{144}C_{F}C_{A}-\frac{11}{36}C_{F}T_{F}n_{f}' = \frac{125}{36} -\frac{11}{54}n_{f}',\label{gamma1}\\
	p_{1}^{\rm NS} &= C_{F}\left(\frac{55}{12}-4\zeta_{3}\right) = \frac{55}{9}-\frac{16}{3}\zeta_{3}\label{p1}.
\end{align}
Here the $n^{\prime}_{f}$-terms in Eq.~\eqref{gamma1} represent the $n_f$-terms that are not related to the $\beta$-function, which cannot be adopted for fixing the correct running behavior of $\alpha_s$. When using the PMC method, these must be treated as conformal terms.

More explicitly, as for the suggested first example of $R(Q)$ with two active flavors at $Q = 1$\,GeV and $\tilde{\Lambda}_{\overline{\rm MS}} = 0.2$ GeV\,\footnote{For $Q = 1$\,GeV, $n_{f}$ should be equal to $3$, since $Q$ is larger than the strange quark mass. However, the conclusion will not change for either choice, so we adopt this suggestion for ease of comparison.}, we have $b=29/6$ and $c = 115/58$ for the first two $\beta$-coefficients. According to Eqs.~(\ref{r0})--(\ref{p1}), the first two coefficients of the initial scale-dependent series for the case of $\mu=Q$ are
\begin{align}
	r_{0} &= 1,\\
	r_{1} &= \frac{365}{24} - 11\zeta_{3} - \frac{11}{72} n_{f}' + \left( -\frac{55}{72} + \frac{2}{3} \zeta_{3} \right) n_{f} \\
	   &= B + A n_{f},
\end{align}
where
\begin{align}
	A &= -\frac{55}{72} + \frac{2}{3}\zeta_{3} = 0.0374824, \\
	B &= \frac{1073}{72} - 11\zeta_{3}    = 1.68015,
\end{align}
which leads to $r_{1} = 2A + B = 1.75512$, and then the NLO conventional series for $\mu=Q$ is
\begin{align}\label{Convi-result}
	R(\mu = Q)|_{\overline{\rm MS}} = a_{\overline{\rm MS}}(Q) + r_{1}(Q) a_{\overline{\rm MS}}^{2}(Q) = 0.0993138,
\end{align}
where $a_{\overline{\rm MS}}(Q)$ is derived from the NLO RGE solution~\eqref{RGEsolution}, e.g.\ $a_{\overline{\rm MS}}(Q)=0.0862557$.

By applying the PMC, the initial series (\ref{convRQ}) will be improved to a scale-independent conformal series:
\begin{align}\label{PMCi-result}
	R_{\rm PMC} = a_{*} + r_{1,0} a_{*}^{2} = 0.0971576,
\end{align}
where the NLO conformal coefficient $r_{1,0}$ is
\begin{displaymath}
r_{1,0} = \frac{33}{2}A + B = 2.29861,
\end{displaymath}
and the PMC scale $\mu^{*}$ is
\begin{displaymath}
\mu^{*} = Q\exp(3A) = 1.11901\,{\rm GeV}.
\end{displaymath}
The magnitude of $a_{*}$ can be numerically derived from Eq.~\eqref{RGEsolution}, e.g.\ $a_{*}=0.0817833$. We point out that in Stevenson's treatment, the $n^{\prime}_{f}$-terms were wrongly adopted for fixing the magnitude of $\alpha_s$, e.g.\ the coefficient $-11/72$ was wrongly put into the coefficient $A$, thus leading to incorrect PMC predictions and discussions. Moreover, in footnote~4 of Ref.~\cite{Stevenson:2023qsh}, Stevenson stated that, ``\emph{This error makes matters worse, in that the BLM result is not even the same as using the $\overline{\rm MS}(\mu = \mu_{r}^{*})$ scheme}.'' Here his meaning of ``$\overline{\rm MS}(\mu = \mu_{r}^{*})$ scheme" is to set $\mu = \mu^{*}$ in the initial $\overline{\rm MS}$-series. Using the wrong BLM/PMC prediction to make the comparison, this comment is incorrect. Moreover, we emphasize that the purpose of PMC is not to find the optimal scale for the pQCD series, but to obtain scale-invariant predictions. Thus, a simple choice of $\mu=\mu^*$ in the initial series generally cannot provide the same precise prediction as that of the PMC. It is interesting to note that for this particular NLO example, by setting $\overline{\rm MS}(\mu = \mu^{*})$ for the initial series, one can obtain the same numerical result as that of the PMC, e.g.\
\begin{align}
 R(\mu = \mu^{*})|_{\overline{\rm MS}} =&\, a_{\overline{\rm MS}}(\mu^{*}) + \left[r_{1}(Q)
 + b \ln\left(\frac{\mu^{*}}{Q}\right)\! \right]\! a_{\overline{\rm MS}}^{2}(\mu^{*})\notag\\
 =&\, 0.0971576,
\end{align}
where the RGE has been implicitly adopted to obtain the full $\mu$-dependence of the series.

\subsection{Comments on the Stevenson's second example}

Using the above-mentioned NLO $e^{+}e^{-}$ total cross-section ratio as the platform, Stevenson's second example concerns the scheme dependence/independence of the PMC predictions. Again, for ease of comparison, we first introduce the same notation as Refs.~\cite{Stevenson:2023isz, Stevenson:2023qsh} to facilitate the calculations and discussions.

Following Refs.~\cite{Stevenson:2023isz, Stevenson:2023qsh}, one can convert a pQCD series from the ${\bf I}$-scheme to another ${\bf II}$-scheme by using the following relation between the strong coupling constants under different schemes\,\footnote{For higher-order pQCD series, one must also transform the expansion coefficients properly via different schemes, so as to obtain the correct pQCD prediction under different schemes.

For example, the correct formulas for transforming a pQCD series among different schemes up to N$^3$LO-level can be found in Ref.~\cite{Ma:2017xef}. Here we accept this simple scheme transformation for the discussion, since it is correct at the presently considered NLO level.}, i.e.
\begin{align}\label{atoa'}
	a_{{\bf I}} = a_{{\bf II}} \left(1+v_{1}^{({\bf I},{\bf II})}a_{{\bf II}}+\cdots \right),
\end{align}
where the NLO coefficient
\begin{displaymath}
v_{1}^{({\bf I},{\bf II})}=v_{10}^{({\bf I},{\bf II})}n_{f} + v_{11}^{({\bf I},{\bf II})}.
\end{displaymath}
Due to the Celmaster--Gonsalves~(CG) relation~\cite{Celmaster:1979dm, Celmaster:1979km}, we have
\begin{align}\label{CG}
	\tilde{\Lambda}_{{\bf I}}=\tilde{\Lambda}_{{\bf II}}\exp(v_{1}^{({\bf I},{\bf II})}/b).
\end{align}
In the following, ${\bf II}$ will be the $\overline{\rm MS}$-scheme, throughout, and ${\bf I}$ represents any other scheme.

For brevity, we use $v_{1}$ to denote $v_{1}^{({\bf I},\overline{\rm MS})}$, and correspondingly, $v_{10}=v_{10}^{({\bf I},\overline{\rm MS})}$ and $v_{11}=v_{11}^{({\bf I},\overline{\rm MS})}$. Thus, we can directly use the $\overline{\rm MS}$-scheme formulas listed in the subsection~\ref{firstexample}. Using the transformation (\ref{atoa'}), one can transform the $\overline{\rm MS}$-series (\ref{convRQ}) into that under the ${\bf I}$-scheme,
\begin{align}
	R(Q) = r_{0}^{{\bf I}} a_{{\bf I}} + r_{1}^{{\bf I}} a^{2}_{{\bf I}} +\cdots,
\end{align}
with coefficients
\begin{align}
	r_{0}^{{\bf I}} &= r_{0} = r_{0,0}=1, \\
	r_{1}^{{\bf I}} &= r_{1} - r_{0} v_{1} = r_{1,0} + r_{1,1}b - r_{0}v_{1},
\end{align}
where the $(-r_{0}v_{1})$-term comes from the transformation of $\alpha_s$ under different schemes, e.g.\ Eq.~(\ref{atoa'}), which should be treated as non-conformal term and be used for fixing the correct magnitude of $\alpha_s$ under the ${\bf I}$-scheme. Following the standard scale-setting procedures, we obtain the PMC series under the ${\bf I}$-scheme
\begin{align}
	R_{\rm PMC} = a_{*} + r_{1,0} a_{*}^{2},
\end{align}
where $ a_{*}= a_{*}(\mu_{\bf I}^{*})$ and
\begin{align}
	r_{1,0}^{{\bf I}} &= r_{1,0} = \frac{33}{2}A + B,\notag\\
	\mu_{\bf I}^{*} &= Q \exp\left( -\frac{r_{1,1}}{r_{0,0}} + \frac{v_{1}}{b} \right).
\end{align}

To discuss the scheme invariance of the PMC predictions, in addition to the $\overline{\rm MS}$-scheme, Stevenson suggested two different schemes with two active flavors, which are characterized by two typical choices of $v_{10}$ and $v_{11}$, e.g.\ the $\bf ii$-scheme, for $v_{10}=0$ and $v_{11}=2$, and the $\bf iii$-scheme, for $v_{10}=1$ and $v_{11}=0$.

For the $\bf ii$-scheme, we have
\begin{align}
	r_{1}^{\bf ii} &= r_{1}-2=-0.244883,\\
	\tilde{\Lambda}_{\bf ii} &= \tilde{\Lambda}_{\overline{\rm MS}}e^{v_{1}/b} = 0.302509\ {\rm GeV},
\end{align}
and the NLO coupling $a_{\bf ii}(Q)$ under the $\bf ii$-scheme can be solved from Eq.~\eqref{RGEsolution} as $a_{\bf ii}(Q)=0.108568$. Then the NLO conventional result under the $\bf ii$-scheme is
\begin{align}
	R(\mu = Q)|_{\bf ii} = a_{\bf ii}(Q) + r_{1}^{\bf ii} a_{\bf ii}^{2}(Q) = 0.105682.
\end{align}
To compare with the result (\ref{Convi-result}) under $\overline{\rm MS}$-scheme, the conventional series is scheme dependent.

After applying the PMC, the corresponding conformal coefficient and scale are
\begin{align}
	r_{1,0}^{\bf ii} &= r_{1,0} = 2.29861,\notag\\
	\mu_{r, \bf ii}^{*} &= 1.69256\ {\rm GeV},
\end{align}
which lead to
\begin{align}\label{PMCii-result}
	R_{\rm PMC}^{\bf ii} = a_{*, {\bf ii}} + r_{1,0}^{\bf ii} a_{*, {\bf ii}}^{2} = 0.0971576,
\end{align}
where the NLO $a_{*,(\bf ii)}$ can be solved from Eq.~\eqref{RGEsolution} as $a_{*,(\bf ii)}=0.0817833$. A comparison of Eqs.~(\ref{PMCi-result}) and (\ref{PMCii-result}) shows that the PMC predictions under both the $\overline{\rm MS}$-scheme and ${\bf ii}$-scheme are the same.

Similarly, for the ${\bf iii}$-scheme, we have
\begin{align}
	r_{1}^{\bf iii} &= r_{1}-2 = -0.244883,\\
	\tilde{\Lambda}_{\bf iii} &= \tilde{\Lambda}_{\overline{\rm MS}}e^{v_{1}/b} = 0.302509\ {\rm GeV},\\
	a_{\bf iii}(Q) &= 0.108568.
\end{align}
Thus, we obtain the same NLO conventional result as that of the $\bf ii$-scheme, i.e.
\begin{align}
	R(\mu = Q)|_{\bf iii} = a_{\bf iii}(Q) + r_{1}^{(\bf iii)} a_{\bf iii}^{2}(Q) = 0.105682.
\end{align}
Accidentally, the predictions of conventional series are the same for $\bf ii$-scheme and $\bf iii$-scheme, both of them are different from the result of the $\overline{\rm MS}$-scheme. After applying the PMC, we have
\begin{align}
	r_{1,0}^{\bf iii} &= r_{1,0} = 2.29861,\notag\\
	\mu_{r, {\bf iii}}^{*} &= 1.69256\ {\rm GeV},
\end{align}
which lead to
\begin{align}\label{PMCiii-result}
	R_{\rm PMC}^{\bf iii} = a_{*, {\bf iii}} + r_{1,0}^{\bf iii} a_{*, {\bf iii}}^{2} = 0.0971576,
\end{align}
where the NLO $a_{*, {\bf iii}}$ can be solved from Eq.~\eqref{RGEsolution} as $a_{*,(\bf iii)}=0.0817833$. A comparison of Eqs.~(\ref{PMCi-result}) and (\ref{PMCiii-result}) shows that the PMC predictions under both the $\overline{\rm MS}$-scheme and ${\bf iii}$-scheme are the same.

Eqs.~(\ref{PMCi-result}), (\ref{PMCii-result}) and (\ref{PMCiii-result}) show that by applying the PMC correctly, one can obtain exactly the same pQCD predictions under different schemes. So Stevenson's conclusion that the PMC cannot solve the renormalization-scheme-dependence problem is incorrect, which is due to his misuse of PMC, e.g. he does not correctly deal with the $(-r_{0}v_{1})$-term so as to set the correct magnitude of $\alpha_s$ and then to achieve the correct PMC series.

As a final remark, in Appendix B of Ref.~\cite{Stevenson:2023isz}, Stevenson claimed,
\begin{quote}
``\emph{We may use the renormalization scheme (RS) choice to achieve exact conformality, with our result for ${\cal R}$ being energy independent. There are several ways to do this. One is to adjust the scheme, decreasing the $\mu/\tilde{\Lambda}$ value, until the coefficient $r_{1}$ becomes so large and negative that $r_{1}a=-1$. The NLO result is then ${\cal R}=a(1-1)=0$. Alternatively, we may make $\mu/\tilde{\Lambda}$ arbitrarily large, and hence $a$ arbitrarily small; the $r_{1}$ coefficient then becomes large and positive, approaching $1/a$ as $a \to 0$; our NLO result is then $a(1+1)=2a\to 0$. We may easily extend these stratagems to higher orders (no actual Feynman-diagram calculations are needed!) to achieve ${\cal R}=0$ to any order}.''
\end{quote}
This argument is evidently incorrect and misleading. Since QCD is a non-conformal theory, one can only approach conformality to the maximum extent via a step-by-step method, as the PMC does. At the NLO level, the magnitude of the strong coupling can be derived from Eq.~\eqref{RGEsolution} and by utilizing the CG relation~\eqref{CG}, we can rewrite Eq.~\eqref{RGEsolution} into proper form for any renormalization scheme that is defined by a different choice of $v_{1}$, i.e.
\begin{align}\label{RGEsolution'}
	\frac{1}{a'}\left(1+ca'\ln\left|\frac{ca'}{1+ca'}\right|\right) =b\ln\left(\frac{\mu}{\tilde{\Lambda}\exp(v_{1}/b)}\right),
\end{align}
where $a'$ is strong coupling under corresponding scheme, $b$ and $c$ are scheme-independent coefficients of the RGE. This indicates that a scheme change at the NLO level is equivalent to a proper scale change. In other words, $a'(\mu)$ in Eq.~\eqref{RGEsolution'} can also be regarded as $a\left(\mu \exp(-v_{1}/b)\right)$.

Moreover, Stevenson's argument, as quoted above, essentially suggests that by selecting a specific and extreme renormalization scale, one could arbitrarily adjust the pQCD prediction to any value, even zero. However, this statement is indistinguishable from describing renormalization scale dependence. Can one truly assert that fixing the renormalization scale to a specific value that sets the pQCD prediction to zero amounts to achieving conformality, even exact conformality ? Clearly, such a claim is untenable!

\end{document}